# Scalable Memory Sharing in Photonic Quantum Memristors for Reservoir Computing

Chaehyeon Lim[1,2†], Hyungchul Park[1†], Beomjoon Chae[1], Jeonghun Kwak[3], Soo-Yeon Lee[3], Namkyoo Park[2§], and Sunkyu Yu[1*]

[1]*Intelligent Wave Systems Laboratory, Department of Electrical and Computer Engineering, Seoul National University, Seoul 08826, Korea*
[2]*Photonic Systems Laboratory, Department of Electrical and Computer Engineering, Seoul National University, Seoul 08826, Korea*
[3]*Department of Electrical and Computer Engineering, Inter-university Semiconductor Research Center, and SOFT Foundry Institute, Seoul National University, Seoul 08826, Korea*

Although photons are robust, room-temperature carriers well suited to quantum machine learning, the absence of photon-photon interactions hinder the realization of memory functionalities that are critical for capturing long-range context. Recently, measurement-based implementations of photonic quantum memristors (PQMRs) have enabled tunable non-Markovian responses. However, their memory remains confined to local elements, in contrast to biological or artificial networks where memory is shared across the system. Here, we propose a scalable PQMR network that enables measurement-based memory sharing. Each memristive node updates its internal state using the history of its own and neighbouring quantum states, thereby realizing distributed memory. By modelling each node as a photonic quantum memtransistor, we demonstrate pronounced enhancements in both classical and quantum hysteresis at the device level, as well as enhanced network-level quantum hysteresis. Implemented as a quantum reservoir, the architecture achieves improved Fashion-MNIST classification accuracy and confidence via increased data separability. Our approach paves the way toward high-capacity quantum machine learning using memristive devices compatible with linear-optical quantum computing.

In harnessing quantum states of light for machine learning [1,2], a critical challenge lies in realizing memory functionalities essential for capturing temporal dependencies and long-range context [3,4]. However, most photonic platforms exhibit temporally memoryless dynamics due to the weakly interacting nature of photons. In parallel with sustained efforts to exploit light-matter interactions via optical nonlinearities [5,6] and phase-change materials [7-9], recent work has demonstrated a measurement-based memristive architecture: the photonic quantum memristor (PQMR) [10-13], which targets process-in-memory operations analogous to those of electronic memristors [14,15]. Sharing a common philosophy with linear-optical quantum computing (LOQC)—where measurement induces effective nonlinear responses [16]—the PQMR further exhibits memristive behaviours by continuously updating the system in response to time-averaged measurement outcomes.

Although the PQMR has shown improved quantum reservoir computing [17-26] and has verified its quantum character through preserved coherence, initial demonstrations have remained limited to localized, device-level operation with respect to memory functionality [10]. In light of coupled synaptic weights [27], coordinated updates across distributed nodes [28], cross-layer information integration in deep learning [29], and the necessity of synchronizing reservoir computing [25], realizing memory sharing across devices becomes critical. Very recently, this question has been partially explored by probing the dynamics of independent local photonic quantum memristors with entangled photon pairs [30]. Nevertheless, given the practical hurdles in generating high-fidelity multiphoton entanglement [31,32], scalable memory sharing across PQMR networks remains an open challenge.

Here, we propose a scalable memory-sharing scheme for PQMRs, facilitating high-capacity quantum reservoir computing. We develop a PQMR network architecture in which each local device updates its state based on the historical evolutions of its own and nearby Fock-state populations. To analyse the memristive behaviour of each local device, we propose a photonic quantum memtransistor (PQMT) model, demonstrating tunable, enhanced classical and quantum hysteresis responses. Alongside the enhancement also in the network-level quantum hysteresis, we apply our design to Fashion-MNIST classification [33] via quantum reservoir computing, achieving more than a twofold improvement in a combined accuracy-confidence metric. Our results via scalable measurement-based feedback can endow improved memory functionalities for quantum machine learning and photonic computing.

*Quantum memtransistor model.* To illustrate the concept of memory sharing in our PQMT networks—the PQMR networks with memory sharing (Fig. 1a)—we review the operation principle of a PQMR [10]: a photonic system with parameters that depend on the evolution of the input quantum state. Figure 1b shows its simplest configuration, consisting of the Mach-Zehnder interferometer possessing the tunable phase shift $\theta$. The input state at port $A$, characterized by $\langle n_A \rangle$, is inferred with the measured $\langle n_D \rangle$ and the system state $\theta(t)$, where $\langle n_X \rangle$ denotes the photon number averaged over a sufficiently short time window at port $X$, which is assumed to be instantaneous in our analysis. Using the obtained $\langle n_A \rangle$ evolution, we update $\theta(t)$ following the previous study [10] (blue lines in Fig. 1a-c) to satisfy

$$T_{\text{PQMR}}(t) = 0.5 + \frac{1}{\tau_{\text{int}}} \int_{t-\tau_{\text{int}}}^{t} \left( 0.5 - \frac{\langle n_A(\tau) \rangle}{\langle n \rangle_{\text{max}}} \right) d\tau, \quad (1)$$

where $T_{PQMR} = \langle n_C \rangle / \langle n_A \rangle$ denotes the non-Markovian transmittance from port $A$ to $C$, $\tau_{int}$ is the characteristic time of memory, and $\langle n \rangle_{max}$ is the maximum mean photon number. Because this update is driven solely by the local input $\langle n_A \rangle$ of each element, the PQMR network—an array of PQMRs—operates with device-localized memory of inputs.

When considering the impact of memory sharing in other fields [27-29], one can envisage the collective sharing of input memory across a network in determining the states of PQMRs. Rather than relying on indirect sharing mediated by quantum features of the input [30], we employ the sharing of measurement-based extractions of input histories across the circuit (red arrows in Figs. 1a and 1c). This approach, which imposes correlation across PQMRs, enables electronics-driven scalability while preserving the measured input features.

To investigate the proposed memory scheme at the device level, we develop the model of a photonic quantum memtransistor (PQMT), in which memory shared from a nearby PQMR is assigned to a measurement along a quantum gate path (from port $E$ to port $F$ in Fig. 1c). Extending Eq. (1), we design the transmission of PQMT, as follows:

$$T_{PQMT}(t) = T_{PQMR}(t) f_{mod}(\langle n_A(\tau) \rangle, \langle n_E(\tau) \rangle; \tau \in [t - \tau_{int}, t], d, p), \quad (2)$$

where $f_{mod} \in [0, 1]$, parameterized by the inflection point $p \in [0, 1]$ and the memory-sharing strength $d \geq 0$, is a sigmoidal gating function determined by the histories of $\langle n_E \rangle$ and $\langle n_A \rangle$ (End Matter A). According to the mathematical form of $f_{mod}$, the critical parameter is $d$: $T_{PQMT}(t) = T_{PQMR}(t)$ when $d = 0$ with the gate-off condition $\langle n_E \rangle = \langle n \rangle_{max}$, whereas larger $d$ with the gate-on, $\langle n_E \rangle < \langle n \rangle_{max}$, leads to $T_{PQMT}(t) < T_{PQMR}(t)$. We note that although port $E$ operates as a classical clock when the input $\rho_{AE}$ over ports $A$ and $E$ is a product state, an entangled input enforces the correlations across the measurement at $D$ and $F$, and the consequent histories determining $\theta$.

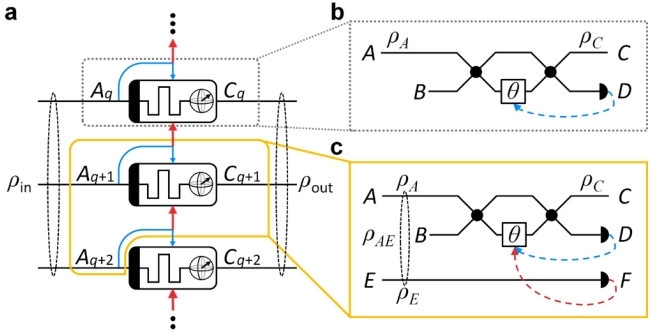

FIG. 1. Memory-shared PQMT network. (**a**) Network architecture. $\rho_{in}$ and $\rho_{out}$ denote the input state over ports $A_q$ and the output state over ports $C_q$, respectively. State update of the $q$-th PQMR is driven locally by feedback from the indirect measurement on port $A_q$ (blue lines) and nonlocally by feedback from the indirect measurement on port $A_{q+1}$ (red lines). (**b**) PQMR model: $\rho_A$ and $\rho_C$ are the states at ports $A$ and $C$. $\langle n_A \rangle$ inferred from the measurement at port $D$ updates $\theta$. (**c**) PQMT model: a gate state $\rho_E$ and the input state $\rho_A$ forms a joint input $\rho_{AE}$, updating $\theta$ through the measurements at $D$ and $F$. The feedback from $F$ corresponds to the memory sharing channel (red lines in (**a**)).

*Device-level hysteresis.* To examine the proposed PQMT operation, we investigate the classical and quantum memristive responses represented by the hysteresis loops of the photon number and quantum coherence (End Matter B), respectively. To evaluate the evolution of these loops in parameter space, we apply a sinusoidally oscillating input state, as follows:

$$|\psi_{AE}\rangle = \left( \cos\left(\frac{\pi t}{\tau_{osc}}\right)|0\rangle + \sin\left(\frac{\pi t}{\tau_{osc}}\right)|1\rangle \right) \otimes \left( \sqrt{1 - \langle n_E \rangle}|0\rangle + \sqrt{\langle n_E \rangle}|1\rangle \right), \quad (3)$$

where $\tau_{osc}$ determines the temporal periodicity.

Figures 2a and 2b demonstrate that memory sharing modelled as the PQMT substantially alters both classical and quantum memristive responses at the device level. In the classical response (Fig. 2a), reducing $\langle n_E \rangle$—which corresponds to stronger memory sharing at a given $d$ according to $f_{mod}$ in Eq. (2)—enhances the hysteresis response in terms of its contrast. Furthermore, despite the decrease of the transmittance induced by gating through $f_{mod}$, the hysteresis of quantum coherence maintains its loop area via increased contrast between the two subloops (Supplementary Note S1 for details). Considering the importance of memory capacity in reservoir computing, especially for time-series tasks [34], the observed enhancement and diversification of memristive responses highlight the advantages of our memory-sharing configuration for quantum reservoir computing.

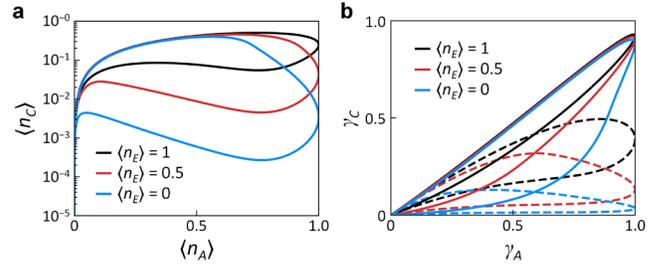

FIG. 2. Hysteresis responses of PQMT. (**a,b**) Hysteresis of $\langle n_C \rangle$ on $\langle n_A \rangle$ (**a**), and of $\gamma_C$ on $\gamma_A$ (**b**) for different values of the gate level: $\langle n_E \rangle = 0, 0.5,$ and $1$, where $\gamma_X$ denotes the quantum coherence at port $X$. Solid and dashed curves in (**b**) correspond to the first and second half-cycles of the input, over the temporal ranges of $m\tau_{osc} < t < (m + 0.5)\tau_{osc}$ and $(m + 0.5)\tau_{osc} < t < (m + 1)\tau_{osc}$, respectively, where $m$ is a cycle index. $\tau_{int}/\tau_{osc} = 0.3$, $\langle n_{max} \rangle = 1$, $p = 0.5$, and $d = 10$.

*Network-level hysteresis.* Extending the device-level analysis, we investigate the network-level performance of memory sharing, focusing on the response of quantum coherence across the PQMT network: $\gamma_{in}$ and $\gamma_{out}$, which denote the multimode output and input coherences, respectively (End Matter B). We implement scalable memory sharing by assembling $N$ PQMR units into a parallel network with cyclic gate coupling—routing the $D_{q+1}$ (or $D_1$) port to the $q$-th (or $N$-th) PQMR to achieve its transmittance update (Fig. 3a). This architecture, which corresponds to an array of PQMTs assigning the nearby $A$-port to the quantum gate path $E$-$F$ in Fig. 1c, couples the cascaded PQMRs through shared history of input evolutions.

To probe the target hysteresis, we apply a time-periodic multimode input given by superpositions of arbitrary two composite Fock states $|i\rangle$ and $|j\rangle$, each with a photon number of at most 2. An example of the $\gamma_{in}$-$\gamma_{out}$ loop is shown in Fig. 3b, achieving the pronounced enhancement of quantum hysteresis for stronger memory sharing with larger $d$. This result demonstrates that the enhancement of quantum memristive behaviours induced by memory sharing, which is observed at the device level (Fig. 2b), is well preserved at the network level as well, which is a critical feature for implementing quantum reservoirs.

This trend is further verified by examining the variation of the hysteresis area, $S_\gamma$, as a function of a memory window $\tau_{int}/\tau_{osc}$ (Fig. 3c). While $S_\gamma$ exhibits oscillatory behaviour with respect to $\tau_{int}/\tau_{osc}$ due to the time-periodic input, it eventually decreases for larger memory windows as a result of averaging over the quantum evolution. Meanwhile, the memory-sharing configuration shows a clear enhancement for $\tau_{int}/\tau_{osc} > 0.5$, demonstrating superior memristive responses over an ensemble of arbitrary multiphoton resources (Supplementary Notes S2 and S3 for extended analysis).

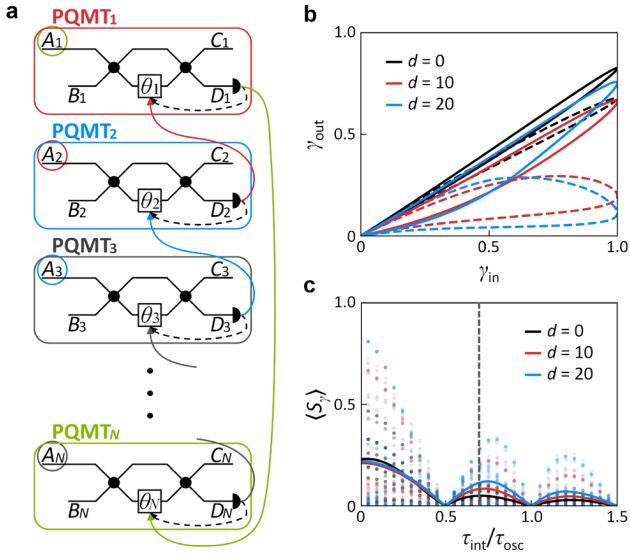

FIG. 3. Memory-shared PQMT network. (**a**) Schematic of $N$ PQMTs connected in parallel with cyclic gate coupling. Photon number at the port $A_q$ is inferred from the measurement on port $D_q$. (**b**) Quantum hysteresis of multimode coherences with varying $d$. The input state is $|\psi_{in}(t)\rangle = \cos(\pi t/\tau_{osc})|0000\rangle + \sin(\pi t/\tau_{osc})|0011\rangle$. Solid and dashed curves correspond to the first and second half-cycles of hysteresis. $\tau_{int}/\tau_{osc} = 0.7$. (**c**) Coherence hysteresis area $\langle S_\gamma \rangle$ as a function of $\tau_{int}/\tau_{osc}$. Solid lines and dots represent the ensemble-averaged result and each realization. The input state realization is defined as $|\psi_{in}(t)\rangle = \cos(\pi t/\tau_{osc})|i\rangle + \sin(\pi t/\tau_{osc})|j\rangle$, where $|i\rangle$ and $|j\rangle$ are the four-mode Fock states with total photon number $\leq 2$. The dashed black line indicates the $\tau_{int}/\tau_{osc}$ used for (**b**). $N = 4$ in (**b**,**c**). Reflecting the decrease of the photon number at each port with increasing $N$, we set the inflection point $p$ following End Matter C.

*Quantum reservoir computing.* Based on the observed enhancements in classical and quantum hysteresis responses, we apply our memory-sharing scheme to quantum reservoir computing (Fig. 4a). Our configuration basically follows a prior protocol [10]; the PQMR (here, PQMT) network is embedded between random-Haar unitary photonic circuits [35] (Fig. 4b). While the unitary circuits facilitate complex linear mappings of sequential data into a high-dimensional Hilbert space, the PQMR network provides the nonlinear and non-Markovian features that are critical for capturing highly complex and long-term dependency [3,4]. We focus on the performance of our PQMT network under the control of the memory sharing strength $d$.

As a fundamental metric for classification tasks, we evaluate the data separability achieved by the PQMTs, quantified by the distance between the transmittance vectors corresponding to an arbitrary pair of Fashion-MNIST images [33] after propagation through the PQMT layer (Fig. 4c). The result demonstrates that the PQMT network with localized memory ($d = 0$, equivalent to the original PQMR) does not fully exploit the vector space spanned by the PQMT transmittance, leading to clustered data representations that usually degrade classification performance [36]. By contrast, enhancing memory sharing with increasing $d$ yields a pronounced expansion of the utilized vector space, corresponding to enhanced discrimination between image patterns.

Figure 4d shows the resulting performance of our quantum reservoir computing, which utilizes the full quantum processor composed of unitary circuits and a memory-shared PQMT network (see End Matter D for details). Performance is evaluated using the averaged confidence and accuracy in the Fashion-MNIST classification task [33], demonstrating superior results in both metrics under the memory-sharing configuration. These results show that, despite employing the simplest form of cyclic ring-type memory sharing (Fig. 3a), our approach achieves a pronounced performance improvement characteristic of a high-capacity reservoir.

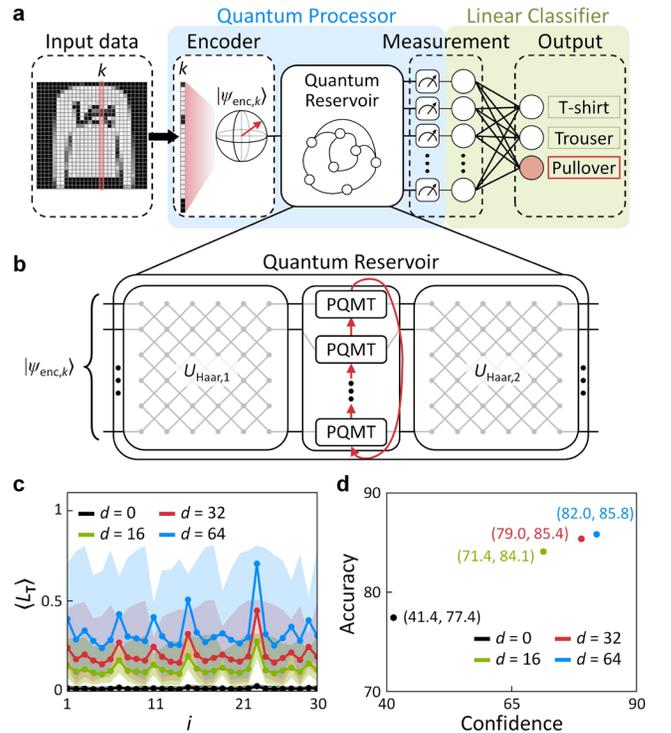

FIG. 4. Quantum reservoir computing with memory-shared PQMT network. (**a**) The entire architecture for image classification. The encoded columns are injected sequentially

into the reservoir. After the processing of final image column, the reservoir output is measured and sent to the linear classifier (see End Matter D for details). (**b**) Reservoir architecture: a parallel PQMT layer of 9 memory-shared PQMTs embedded between two random Haar unitary circuits ($U_{\text{Haar},1}$ and $U_{\text{Haar},2}$). (**c**) Separability of PQMT network over 30 input images, quantified by the ensemble average $\langle L_T \rangle = \sum_j \|\mathbf{T}_i - \mathbf{T}_j\|_2 / 30$ for the $i$-th image, where $\mathbf{T}_i = [T_{\text{PQMT},1}, T_{\text{PQMT},2}, \ldots, T_{\text{PQMT},9}]$ denotes the final transmittance vector after processing the $i$-th image, and $\|\cdot\|_2$ is the L2-norm. The shaded regions indicate the ranges from the minimum to the maximum of realizations. (**d**) Accuracy and confidence of label prediction on varying $d$, evaluated for task of three class classification of Fashion-MNIST dataset.

In the context of the proposed memory-sharing scheme, which enables long-range coupling, additional degrees of freedom remain to be explored for further performance enhancement in machine learning tasks. First, while we focus on the simplest memory-sharing architecture—a regular, directed network with node degree two—we can envisage more complex memory-sharing networks, including higher node degrees, bidirectional couplings, random graphs, or even small-world architectures [37]. Because the optimal network architecture may be task-specific, systematic studies across a broader range of machine learning tasks are expected.

In addition, considering recent advances in modal mixing within linear optics—such as waveguide arrays implementing discrete fractional Fourier transforms [38,39] and synthetic-dimensional matrix calculations [40,41]—it is possible to incorporate elements that simultaneously provide linear optical functionality and modal mixing. Such configurations may offer an optimal platform for memory sharing, as each device port can already encode information arising from light-matter interactions distributed across the entire structure.

In summary, we have proposed and demonstrated a scalable, measurement-based global memory-sharing scheme among PQMTs. By incorporating the historical evolution of both local and neighbouring Fock-state populations across devices, the proposed scheme extends localized non-Markovian dynamics into network-level, correlated memory. Both device-level analysis based on the PQMT model and network-level investigations reveal tunable and enhanced hysteresis responses. Leveraging the resulting memristive nonlinearity and long-term temporal dependence, we have applied our architecture to Fashion-MNIST classification within a quantum reservoir computing framework, achieving more than a twofold improvement in the combined accuracy-confidence metric compared with conventional PQMR networks. The observed improvements in class separability and classification accuracy—beyond this simplest example of cyclic memory sharing—establish scalable, measurement-based configurations as a practical route toward strengthened memory functionalities for quantum machine learning [1,2] and photonic computing [42].

We acknowledge financial support from the National Research Foundation of Korea (NRF) through the Basic Research Laboratory (No. RS-2024-00397664), Innovation Research Center (No. RS-2024-00413957), Pilot and Feasibility Grants (No. RS-2025-19912971), and Midcareer Researcher Program (No. RS-2023-00274348), all funded by the Korean government. This work was supported by Creative-Pioneering Researchers Program and the BK21 FOUR program of the Education and Research Program for Future ICT Pioneers in 2025, through Seoul National University. We also acknowledge an administrative support from SOFT foundry institute.

§nkpark@snu.ac.kr
*sunkyu.yu@snu.ac.kr

**End Matter**

**A. PQMT operation**

We define the PQMT transmission response via the gating function $f_{\text{mod}}$, as follows:

$$f_{\text{mod}}\left(\langle n_A(\tau)\rangle,\langle n_E(\tau)\rangle;\tau\in[t-\tau_{\text{int}},t]\right) \\ = G_{p,d}\left(H\left(\langle n_A(\tau)\rangle,\langle n_E(\tau)\rangle;\tau\in[t-\tau_{\text{int}},t]\right)\right), \quad (A1)$$

where $H$ denotes the history mapping function:

$$H\left(\langle n_A(\tau)\rangle,\langle n_E(\tau)\rangle;\tau\in[t-\tau_{\text{int}},t]\right) \\ = 0.5 + \frac{1}{\tau_{\text{int}}}\int_{t-\tau_{\text{int}}}^{t}\left(0.5 - \sqrt{\frac{\langle n_A(\tau)\rangle}{\langle n\rangle_{\max}}\left(1-\frac{\langle n_E(\tau)\rangle}{\langle n\rangle_{\max}}\right)}\right)d\tau, \quad (A2)$$

$G_{p,d}(x)$ is the sigmoidal nonlinearity function, which is parameterized by the inflection point $p$, the memory-sharing strength $d$, and the other hyperparameters $a_{1-3}$:

$$G_{p,d}(x) = \begin{cases} a_1 e^{dx} & (x < p) \\ a_2 - a_3 e^{-dx} & (x \geq p) \end{cases}. \quad (A3)$$

We set the normalization condition as $G_{p,d}(1) = 1$, and the continuity and differentiability at the $p$, leading to

$$a_1 = \frac{1}{2e^{dp} - e^{2dp-d}}, \quad a_2 = 2e^{dp}a_1, \quad a_3 = e^{2dp}a_1. \quad (A4)$$

**B. L1-norm quantum coherence**

The quantum coherence on a single bosonic mode with the Fock basis is defined as:

$$\gamma_X = \sum_{n\neq m}|\langle n|\rho_X|m\rangle|, \quad (B1)$$

The coherence of an $N$-mode bosonic state with respect to the multimode Fock basis is

$$\gamma_{X_1,\cdots X_N} = \sum_{\mathbf{n}\neq\mathbf{m}}|\langle\mathbf{n}|\rho_{X_1,\cdots X_N}|\mathbf{m}\rangle| \quad \left(|\mathbf{n}\rangle\equiv|n_1,\cdots,n_N\rangle_{X_1,\cdots X_N}\right), \quad (B2)$$

where $(X_1, X_2, \ldots, X_N)$ represent a set of ports in our analysis.

**C. Inflection point setting**

In the PQMT network analysis described in Figs. 3 and 4, we set the inflection point $p$, as follows:

$$p = 1 - \sqrt{\frac{1}{N}\left(1-\frac{1}{N}\right)}, \quad (C1)$$

to reflect the decrease of the port-wise photon number with increasing $N$.

**D. Reservoir computing**

**D1. Quantum reservoir architecture**

The PQMT quantum reservoir (Fig. 4b) consists of two linear mixing stages described by the random Haar unitary circuits designed using QR decomposition. The reservoir comprises $N = 9$ memory-shared PQMT units. The inflection point $p$ is set according to End Matter C. The simulations are performed with the input photon number bound $n \leq 3$ and the update parameter $\langle n\rangle_{\max} = 3$.

**D2. Data encoding and training protocol**

Each Fashion-MNIST image has the resolution of 28×28 pixels. Let $a_{j,k} \in [0, 1]$ denote the normalized grayscale value of pixel $(j, k)$. Each column $k$ is encoded into a quantum input state as:

$$|\psi_{\text{enc},k}\rangle = \sum_{j=1}^{28}(a_{j,k}+\varepsilon)|j\rangle \quad (D1)$$

where $|j\rangle$ is a 9-mode Fock basis with total photon number $1 \leq n \leq 3$, and $\varepsilon = 10^{-8}$ ensures well-defined encoding for fully black columns (i.e., $\forall j, a_{j,k} = 0$). After processing all 28 columns, we record the PQMT transmittance vectors for Fig. 4c. We then re-inject the final column state and measure the output probability distribution over photon-arrival events on the measurement layer after the second unitary mixing layer. The measurement is assumed to be photon number resolving, that is, able to distinguish distinct arrival number outcomes.

The resulting output probability vector is supplied to a linear classifier (single linear layer followed by the softmax function). The classifier is trained for 100 epochs using the Adam optimizer with the learning rate 0.05 and the batch size 256. The reported accuracy and confidence (Fig. 4d) are obtained by averaging metrics over the final five epochs. The confidence score is defined as the dataset average of the maximum softmax probability, defined as:

$$\text{Confidence} = \frac{1}{D_T}\sum_{i=1}^{D_T}\max_c p(y_i = c|x_i) \quad (D2)$$

where $x_i$ and $y_i$ denote the $i$-th input image and its label, respectively, $D_T = 3000$ is the test dataset size, and $c$ is the class label.

**Supplementary Information for "Scalable Memory Sharing in Photonic Quantum Memristors for Reservoir Computing"**

Chaehyeon Lim[1,2†], Hyungchul Park[1†], Beomjoon Chae[1], Jeonghun Kwak[3], Soo-Yeon Lee[3], Namkyoo Park[2§], and Sunkyu Yu[1*]

[1]Intelligent Wave Systems Laboratory, Department of Electrical and Computer Engineering, Seoul National University, Seoul 08826, Korea

[2]Photonic Systems Laboratory, Department of Electrical and Computer Engineering, Seoul National University, Seoul 08826, Korea

[3]Department of Electrical and Computer Engineering, Inter-university Semiconductor Research Center, and SOFT Foundry Institute, Seoul National University, Seoul 08826, Korea

E-mail address for correspondence: [§]nkpark@snu.ac.kr (N.P.), [*]sunkyu.yu@snu.ac.kr (S.Y.)

**Note S1. Device-level hysteresis**

**Note S2. Network-level hysteresis: Varying $N$**

**Note S3. Network-level hysteresis: Input dependence**

**Note S1. Device-level hysteresis**

We provide details of the analysis of the input-output relation for a PQMT. For Fig. 1c of the main text, the input state at port $A$, $|\psi_A\rangle = (\alpha|0\rangle + \beta|1\rangle)$, evolves into $|\psi_{CD}\rangle = \alpha|0\rangle|0\rangle + \beta(\cos(\theta/2)|1\rangle|0\rangle + i\sin(\theta/2)|0\rangle|1\rangle)$ at ports $C$ and $D$, where each beam splitter in the interferometer is implemented using two 50:50 directional couplers and a $\pi/2$ phase shifter along the B-D path. Projecting the output onto port $C$ yields $|\psi_C\rangle = \alpha|0\rangle + \beta(\cos(\theta/2)|1\rangle + i\sin(\theta/2)|0\rangle)$. Using the L1-norm definition of coherence (End Matter B), the coherences of the states at port $A$ and port $C$ are evaluated as:

$$\gamma_A = 2|\alpha\beta|, \quad \gamma_C = \left(\alpha^*\beta + \alpha\beta^*\right)\cos(\theta/2), \tag{S1}$$

respectively, where $\gamma_X$ denotes the quantum coherence of the state at port $X$.

To quantify hysteresis, we use input state in Eq. (3) of the main text, computing the area enclosed by the input-output hysteresis curves. The resulting hysteresis loop areas of the photon number ($S_n$) and coherence ($S_\gamma$) are shown in Fig. S1a and S1b, respectively. Because the overall transmission and output coherence decreases in the gate-on state, $\langle n_E \rangle = 0$, the classical hysteresis loop area $S_n$ is reduced. However, the contrast between upper and lower branches becomes more pronounced (Fig. 2 of the main text). Moreover, the quantum hysteresis area increases markedly (Fig. S1b), highlighting the potential of the PQMT as a building block for quantum reservoirs.

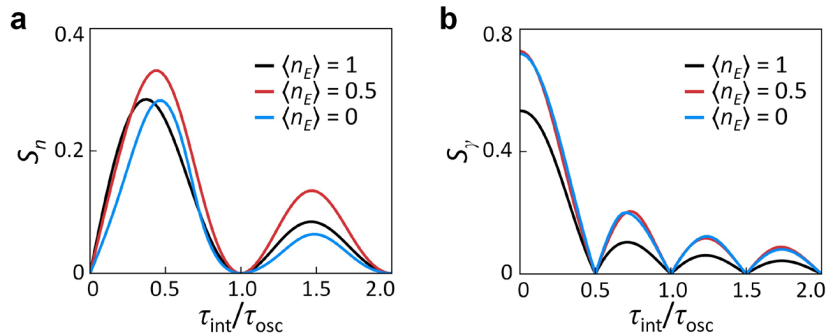

**Fig. S1.** Device-level hysteresis in a PQMT. (**a,b**) Areas of the photon number hysteresis $S_n$ (**a**) and the coherence hysteresis $S_\gamma$ (**b**) on characteristic memory time $\tau_{int}/\tau_{osc}$, respectively.

**Note S2. Network-level hysteresis: Varying $N$**

We compare the network-level hysteresis of the PQMT array for different network sizes $N$. Figures S2a and S2b show the coherence hysteresis area $\langle S_\gamma \rangle$ evaluated for $N = 2$ and $N = 3$ networks, respectively, using the same method employed as in Fig. 3c of the main manuscript. An increase in $\langle S_\gamma \rangle$ is observed for both $N = 2$ and $N = 3$ as $d$ increases, indicating an enhanced hysteresis response independent of system size.

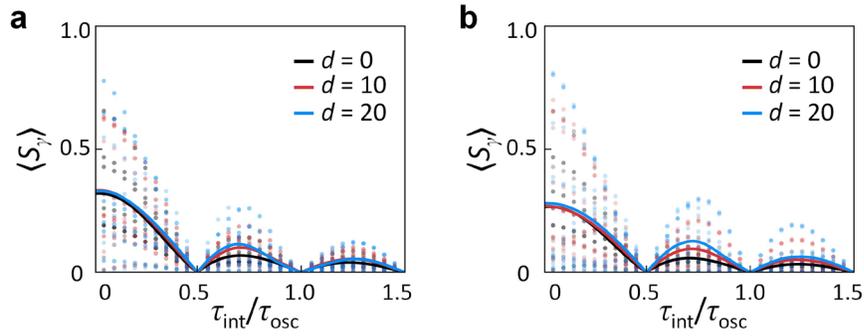

**Fig. S2.** Network-level hysteresis for $N = 2$ (**a**) and $N = 3$ (**b**). All the other parameters and the applied methods are the same as those in Fig. 3c of the main text.

**Note S3. Network-level hysteresis: Input dependence**

We examine how the hysteresis area depends on the choice of basis used to define the input state. As in Fig. 3, the input is set to be the sinusoidally oscillating state $|\psi_{\rm in}(t)\rangle = \cos(\pi t/\tau_{\rm osc})|i\rangle + \sin(\pi t/\tau_{\rm osc})|j\rangle$. For a given basis state $|i\rangle$, we define the hysteresis area, as

$$\langle S_\gamma \rangle^{|i\rangle} \equiv \langle S_\gamma(|i\rangle, |j\rangle) \rangle_{j \neq i}, \tag{S2}$$

which corresponds to the hysteresis area averaged over all possible Fock states $|j\rangle$. This quantity serves as a figure of merit for selecting basis $|i\rangle$ that yields a larger hysteresis area. In our analysis, the network is invariant under cyclic permutations of the mode indices, and therefore, $S_\gamma$ exhibits the same invariance under such transformations.

The resulting state dependence (Fig. S3a) indicates that highly mode-localized photon distributions, such as $|0002\rangle$, are less effective at producing large $S_\gamma$ in the PQMT network of $d = 10$. By contrast, this state dependence is less pronounced in the conventional PQMR network with $d = 0$ (Fig. S3b).

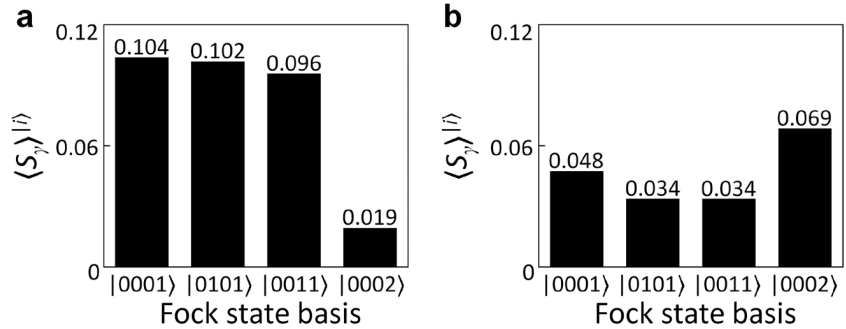

**FIG. S3.** Input state dependence of network hysteresis. Marginal hysteresis $\langle S_\gamma \rangle^{|i\rangle}$ for **(a)** $d = 10$ and **(b)** $d = 0$. Vacuum state is excluded. All the other parameters are the same as those in Fig. 3c.